\documentclass[%
reprint,
 amsmath,amssymb,
pra,
]{revtex4-2}

\usepackage{graphicx}
\usepackage{dcolumn}
\usepackage{bm}

\begin{document}

\title{Topological edge states of Kekul\'e-type photonic crystals induced by a synchronized rotation of unit cells}

\author{{Rui Zhou}\textsuperscript{1}}

\author{{Hai Lin}\textsuperscript{1}}
\email{Corresponding author: linhai@mail.ccnu.edu.cn}

\author{{Y. Liu}\textsuperscript{2}}
\email{Corresponding author: yangjie@hubu.edu.cn}
\author{{Xintong Shi}\textsuperscript{1}}
\author{{Rongxin Tang}\textsuperscript{3}}
 
\author{{Yanjie Wu}\textsuperscript{1}}

\author{{Zihao Yu}\textsuperscript{1}}

\affiliation{$^{1}$ College of Physics Science and Technology, Central China Normal University, Wuhan 430079, Hubei Province \\
$^{2}$School of Physics and Electronic Sciences, Hubei University, Wuhan 430062, Hubei Province\\
$^{3}$Institute of Space Science and Technology, Nanchang University, Nanchang 330021, Jiangxi Province}%

\date{submitted to Phys. Rev. A (Lett.) on 12nd Apr, 
revised 06th Jul., accepted 10th Aug., proof-read 13rd Sept. 2021.  }

 \begin{abstract}
Generating and manipulating Dirac points in artificial atomic crystals has received attention especially in photonic systems due to their ease of implementation. In this paper, we propose a two-dimensional photonic crystal made of a Kekul\'e lattice of pure dielectrics, where the internal rotation of cylindrical pillars induces optical Dirac-degeneracy breaking. Our calculated dispersion reveals that the synchronized rotation reverses bands and switches parity as well so as to induce a topological phase transition. Our simulation demonstrates that such topologically protected edge states can achieve robust transmission in defect waveguides under deformation, and therefore provides a pragmatically tunable scheme to achieve reconfigurable topological phases. 
\end{abstract}

\maketitle

\section{\label{sec:level1}
Introduction}

In the past decades, topological photonics has become a rapidly developing research field that aims to explore the wave physics of topological phases of matter in analogue. The concept of topology insulators in condensed matter physics has been lent to many wave physics fields~\cite{2005Quantum,Fu2011Topological}. Haldane and Raghu~\cite{2008Possible,Haldane1988Model} first brought the quantum Hall effect (QHE) to the field of photonics and theoretically proved the photonic quantum Hall effect (PQHE), which opened new avenues for topologically protected optical transmission devices (such as topological lasers, waveguides, and quantum circuits)~\cite{Khanikaev2012Photonic,2014Topological}. The topological edge states generated on the interface between different topological phases promise fascinating features such as robust transmission, backscattering suppression and defect immunity. Topological photonic devices enabling edge state transmission have brought unprecedented opportunities in controlling the electromagnetic (EM) waves at the microwave and optical frequency bands~\cite{2005Extrinsic,2008Possible, Khanikaev2012Photonic}. 

The first realization of a photonic topological state was based on the microwave platform of gyromagnetic photonic crystals by applying a magnetic field to break the time-reversal (TR) symmetry of the system~\cite{WangZ2008, WangZ2009}. However, a weak gyromagnetic effect hinders their extension towards higher-frequency bands and more manipulative scenarios. To sidestep this issue, all-dielectric photonic crystals (PCs) with judiciously designed unit cells have been proposed~\cite{2008Possible, Miyake2016Design} to achieve topological phases through adjusting geometrical structures in the primitive cell of the all-dielectric PCs with $C_6$ symmetry~\cite{HuangDirac, Xu2020Pulse, 2017Multiple, 2016Accidental}. Floquet insulators, the valley Hall effect and the photonic quantum spin Hall effect (PQSHE) have been realized both in theory and in experiment~\cite{Xu2010Quantum,2008Possible,Wang2020Valley,RN2079}. The PCs with $C_6$ symmetry including circular pillars (dielectric cylinders or pores), core shells (circular rings), and elliptical dielectric cylinder clusters were proposed to achieve quantum spin Hall (QSH) effects~\cite{2008Possible,Xu2010Quantum, Wu2015Scheme}. At the interface between different topological phases, the helical edge states are sandwiched which lead to unidirectional nonscattering propagation. This unique robust feature can be used to realize band-gap devices such as topology lasers, integrated optical circuits, etc.~\cite{Xu2010Quantum,2020Topological,2020Recent,Jung2018Midinfrared}. 

In this paper, we propose an internal rotating mechanism to achieve topological edge states via sandwiching two distinct topological phases, on a two-dimensional PC made of a Kekul\'e lattice in dielectrics~\cite{Xu2010Quantum,2008Possible, 2017Acoustic, Wang2020Valley}. This proposal provides a flexible way to achieve optical topological phase transitions. The physical principle is, by the internal rotation, that a crystal lattice with $C_{6v}$ symmetry (representing six fold rotational symmetry and mirror symmetry in six different directions) turns to $C_6$~\cite{2010The}. Destruction of the mirror symmetry of the lattice causes the Dirac point $\Gamma$ to break its degeneracy. The generation and breaking of the Dirac degeneracy result in the occurrence of topological phase transitions~\cite{2017Acoustic}. At the interface of different topological phases generated in this manner, we observe in simulation the edge states protected by the pseudo time-reversal (TR)~\cite{2010The,Chang2013Experimental} symmetry accompanied by $C_6$ symmetry. As a merit of that, we also design a bent waveguide and a defected one and observe in simulation that even in the presence of sharp turns or defects, these edge states still propagate robustly.  

\section{\label{sec:level2} Theory and model}
 
The structure of the proposed artificial meta-atom in a Kekul\'e lattice is shown in Fig.~1(a), where the black solid hexagons embedding six cylindrical pillars in dark blue are the original units of a Kekul\'e lattice. An odd parity for spatial inversion exists at the $\Gamma$ point of the Kekul\'e unit, and also double-degenerate Dirac cones occur~\cite{2008Possible, 2010The}. Therefore, with the hexagon in composition of the triangular lattice, $a_{0}$ is the lattice constant and $\mathbf{a}_{1}$ and $\mathbf{a}_{2}$ are unit vectors. The geometric parameters used in the model throughout this paper are ${a}_{0}=1, {a}_{0}/R=2.92, d=0.11{a}_{0}$ unless otherwise stated. The outstanding pillars are made of yttrium iron garnet (YIG) material ($\varepsilon_d =11.7$) and the background media is set as air. In the enlarged view in Fig.~1(b) of the hexagonal cluster, $R$ is the length from the center of the hexagon to the center of the cylinder, and $d$ is the diameter of the cylindrical pillars. For the rotation mechanism, Fig.~1(c) shows the diagram of the supercell when the cylindrical pillars rotate simultaneously for an angle of $\alpha=9.4^\circ$, clockwise around the center of the hexagon. 

\begin{figure}
\includegraphics[width=0.5\textwidth]{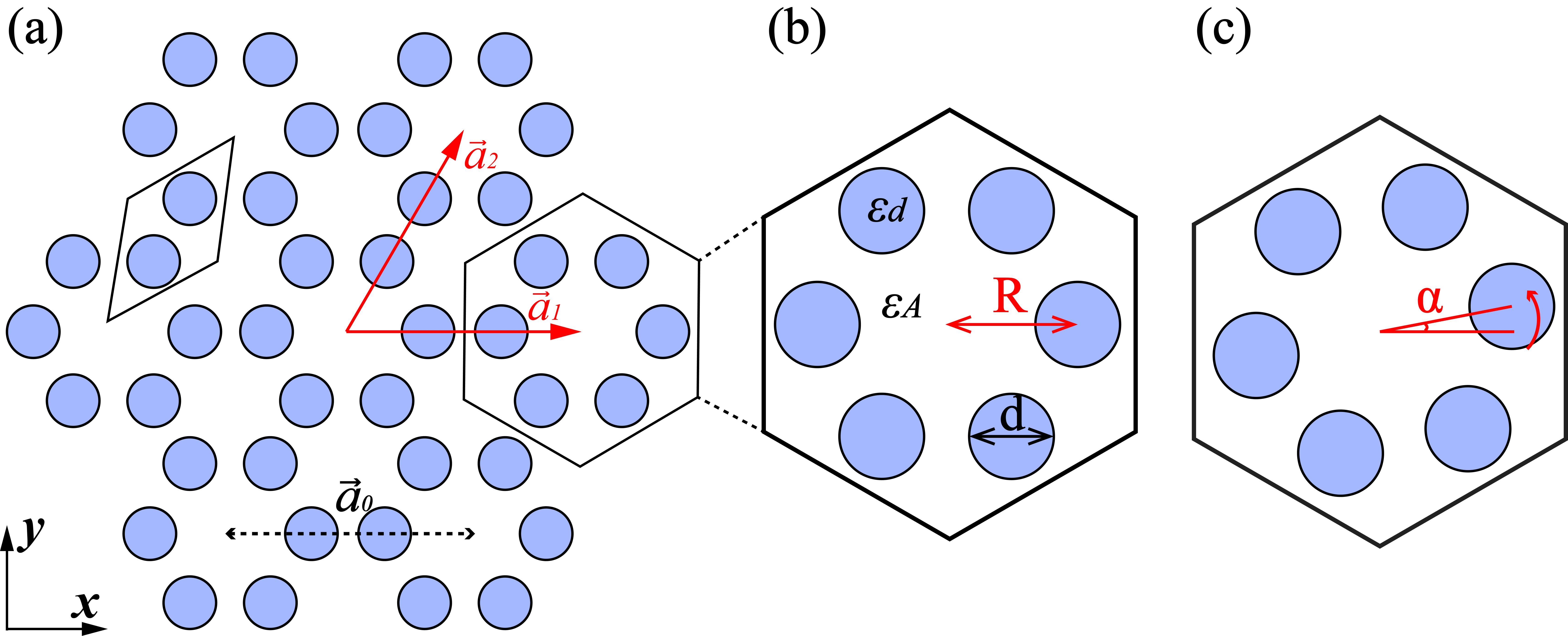}
\caption{\label{fig:epsart}Schematic diagram of a two-dimensional(2D) Kekul\'e lattice with a lattice constant of ${a}_{0}$. (a) The arrangement of the lattice unit marked by the black solid line and the lattice red vectors. (b) Unit cell before rotation. (c) Unit cell under a rotation angle $\alpha=9.4^{\circ}$. }
\end{figure}

In order to facilitate the description of the topological phase transition mechanism, we consider the behavior of the transverse magnetic (TM) mode in a Kekul\'e PC with $C_6$ symmetry~\cite{WangZ2008}. According to Maxwell's equation, the propagation of time-harmonic TM waves in PC can be described by \cite{2008Possible}: 
\begin{equation}
\left[\frac{1}{\varepsilon(\mathbf{r})} \nabla \times \nabla \times\right] E_{z}(\mathbf{r}) \hat{z}=\frac{\omega^{2}}{c^{2}} E_{z}(\mathbf{r}) \hat{z},
\end{equation}
where $\varepsilon(\mathbf{r})$ is the dielectric parameter and $c$ the speed of light. When $\varepsilon(\mathbf{r})$ is periodic, the Bloch theorem applies in the form of EM waves.

For the representations of $\mathbf{E}_{z}$ field at $\Gamma$ point, artificial atoms carry orbitals $p_{x} (p_{y})$ and $d_{x y} (d_{x^{2}-y^{2}})$, in analogy to the electron orbitals for a periodic array of atoms in solid. Since a direct counterpart of the spins does not occur naturally, two eigenvectors $\mathbf{E}_{1}$ and $\mathbf{E}_{2}$ can be constructed as $[p_{+}, p_{-}]$  and $[d_{+}, d_{-}]$ (cf. Sec.~$\mathrm{I}$ of Supplemental Material~\cite{SM}), in which 
 \begin{equation}
p_{\pm}=\frac{p_{x}\pm i p_{y}}{\sqrt{2}}, d_{\pm}=\frac{d_{x^{2}-y^{2}} \pm i d_{x y}}{\sqrt{2}}. 
\end{equation}
Here, the antiunitary operator $T=UK$ is proposed where $U=-i\sigma_y$, and $K$ is a complex conjugate operator~\cite{2008Possible,Xu2010Quantum,Wu2015Scheme}. Since $U^{2}=-1$  guarantees $T^{2}=-1$, $T$ can be used as a pseudo-TR operator in our photonic system. Under the action of $T$ operator, $[p_{+}, p_{-}]$ has the following transformations~\cite{Wu2015Scheme}:
\begin{equation}
T^{2} p_{\pm}=-p_{\pm}. 
\end{equation}
Obviously, this pseudo-TR resulting from the crystal symmetry~\cite{2010The,He2016Photonic,2008Possible} plays the central role in our analogue QSHE. In other words, since pseudo-TR symmetry and pseudo-spin depend on $C_6$ symmetry, photons with pseudo-spins in the $C_6$  system will produce Kramer's degeneray. It is worth clarifying that $C_6$ plus TR symmetry is a true combination of protection degeneracy, and the TR operator is not the main basis for providing the Kramer's degeneracy. 

In an analogue electronic system~\cite{2010The}, the two eigenstates of odd parity in the photonic system can be represented by $p_{+}$ and $p_{-}$ corresponding to the $p$-band pseudo-spin-up and pseudo-spin-down states. Similarly, the two states of even parity in $d_{\pm}$, can be mapped as the pseudo-spin-up and pseudo-spin-down states for $d$-band. 

According to the $\mathbf{k}\cdot \mathbf{p}$ theory~\cite{1989P}, an effective photonic Hamiltonian under the representation of $\left[\begin{array}{llll} p_{+}, & d_{+}, & p_{-}, & d_{-}\end{array}\right]^{T}$ is (cf. Sec.~$\mathrm{II}$ of Supplemental Material~\cite{SM})
\begin{equation}
H_{0}=\left[\begin{array}{cccc}
\omega_{\mathrm{p}}^{2} / c^{2} & A k_{+} & 0 & 0 \\
A^{*} k_{-} & \omega_{\mathrm{d}}^{2} / c^{2} & 0 & 0 \\
0 & 0 & \omega_{\mathrm{p}}^{2} / c^{2} & A^{*} k_{-} \\
0 & 0 & A k_{+} & \omega_{\mathrm{d}}^{2} / c^{2}
\end{array}\right] 
\end{equation}
In Eq.~(4), $k_{\pm}=k_{x} \pm \mathrm{i} k_{y}$, $A$ is the coupling coefficient between the $p$- and $d$-states, and $\omega_{\mathrm{p}}$ and $\omega_{\mathrm{d}}$ the eigenfrequencies of the $p$-band and $d$-band, respectively. Note that in a $C_6$ symmetric system, only $p$-states ($d$-states) with the same spin direction can be coupled and Eq.~(4) is similar to the electronic Hamiltonian in the Bernevig-Hughes-Zhang (BHZ) model~\cite{Bernevig2006, Zhu2014Floquet, Juergens2014Screening}, where the two block matrices correspond to the massive Dirac equations with pseudo-spin-up and down, respectively. This pair of pseudo-spin pairs are interconnected by an inversion symmetry operation, along with their disparate parities which ensures that the entire system satisfies pseudo TR symmetry. If the $p$-band is compared to the valence band and the $d$-band to the conduction band~[cf.~Fig.~2(c)] analogous to the BHZ model in the electronic system, Eq.~(4) serves as a Hamiltonian matrix of PQSHE. Then the topological state of our system should be determined as follows~\cite{Chang2013Experimental}. When $\omega_{\mathrm{p}}>\omega_{\mathrm{d}}$, the system corresponds to topological non-trivial states; when~$\omega_{\mathrm{p}}<\omega_{\mathrm{d}}$, the system has parity inversion at the $\Gamma$ point, which directly indicates the topological trivial state; when $\omega_{\mathrm{p}}=\omega_{\mathrm{d}}$, at the $\Gamma$ point there occurs a band of fourfold degeneracy, corresponding to the double Dirac point which marks the transition point of topological phases~\cite{HuangDirac}. 
\begin{figure}
\includegraphics[width=0.5\textwidth]{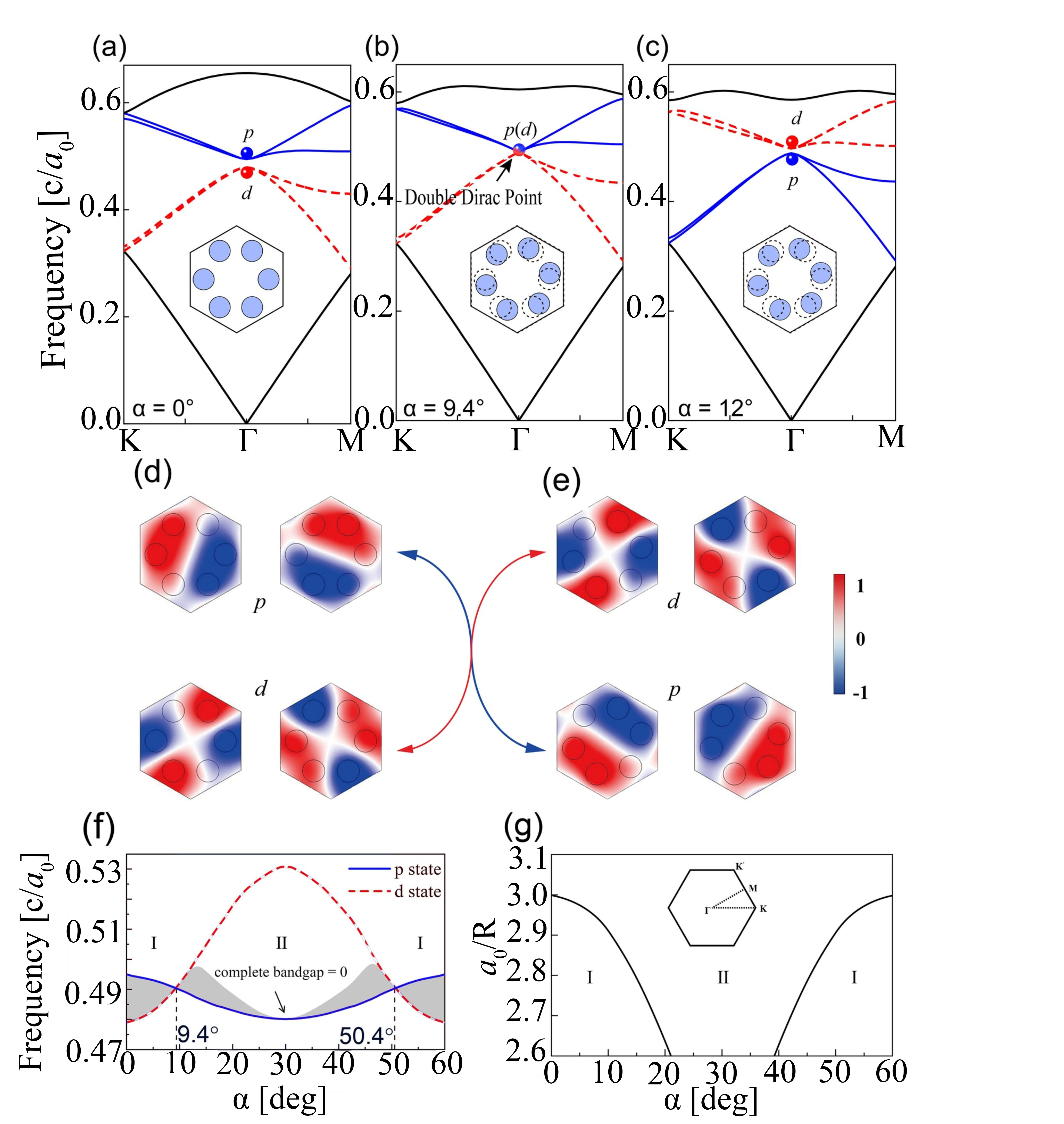}
\caption{\label{fig:epsart} TM mode dispersion diagrams (a-c) and eigenstate distribution (d) when the PC unit cells rotate for different angles. (a) Rotation angle $\alpha=0^{\circ}$. (b) $\alpha=9.4^{\circ}$ where band inversion (red dotted line and  blue solid line represent the d-band and p-band) occurs to generate an accidental Dirac point.  (c) $\alpha=12^{\circ}$. (d) $\mathbf{E}_{z}$ field of the p-band and d-band in (a). (e) $\mathbf{E}_{z}$ field of the d-band and p-band in (c). The black dashed lines in (b-c) indicate the unrotated unit cell. (f) Working frequency for rotation angle $\alpha$ with $a_0/R=2.92$. The two regions $\mathrm{I}$ and $\mathrm{II}$ represent two different topological states of PC: $\mathrm{I}$ is the topological nontrivial phase, and $\mathrm{II}$ the trivial one. The shaded region represent the bandgap width. (g) Phase diagram with rotation angle $\alpha$ and lattice constant $a_0/R$ (inset: the first Brillouin zone of the Kekul\'e lattice). } 
\end{figure}

\section{RESULTS AND DISCUSSION}
 
In this work, \texttt{COMSOL} software based on the finite element method is used to calculate the PC dispersion and the electric field diagram. Considering the TM mode ($\mathbf{E}_{z}$, $\mathbf{H}_{x}$ and $\mathbf{H}_{y}$ components only), the band degeneracy and its breaking at $\Gamma$ point are achieved by adjusting the rotation angle of hexagons in every unit. As show in Fig.~2, when the rotation angle is $12^\circ$, the two degenerate bands split in proximity of $\Gamma$ point. 

Based on the above theory, two degenerate band eigenstates are analogous to the quantum electron wave functions for the $p$-band (blue solid line) and $d$-band (red dotted line)~\cite{2008Possible, Dixon1977Construction}. According to $\mathbf{E}_{z}$ fields in Fig.~2(d) and (e), we recognize that the dipole electric field belongs to the p-band and the quadrupole one to the d band~\cite{Yang2004Dipole}. Therefore, the type of energy band can be determined according to $\mathbf{E}_{z}$ fields. In Fig.~2(a), when the pillars are unrotated $\alpha=0$, there are two degenerate points in each band. By observing the eigenfrequencies of $\mathbf{E}_{z}$ fields,  we find that the p-band goes above the d-band across the band gap. Here in panel (a) our system is initially in a topological nontrivial state ($\omega_{\mathrm{p}}>\omega_{\mathrm{d}}$, represented by phase $\mathrm{I}$). When the unit cell is rotated counterclockwise by $9.4^\circ$ ($\alpha=9.4^\circ$), the eigenfrequency $\omega_{\mathrm{p}}=\omega_{\mathrm{d}}$ at Dirac point $\Gamma$ in panel (b)~\cite{HuangDirac}. When we continue to adjust the rotation angle to $\alpha=12^\circ$, as shown in (c), the fourfold degeneracy point at $\Gamma$ point reopens. We then realize the band degeneracy and its breaking via a simple rotation of the unit cell. Panel (c) also indicates that p band goes under d band in contrast with (a) where the p and d bands are reversed at $\Gamma$ point, reducing it to a topologically trivial state ($\omega_{\mathrm{p}}<\omega_{\mathrm{d}}$, represented by phase $\mathrm{II}$). The physical reason for topological phase switching results both from the band inversion due to broken spatial inversion symmetry and also from the parity inversion~\cite{2010The,Haldane1988Model}. 

The eigenfrequencies of the p- and d- bands within a period for $60^\circ$ changes along with rotation angle $\alpha$, as shown in Fig.~2(f). We find that the p- and d- bands reverse twice in one period of $\alpha$, which occurs at both $\alpha=9.4^\circ$ and $50.4^\circ$ as the topological phase transition points. Moreover, the complete band gap produced by the rotation mechanism increases to peak at $\alpha=13^\circ$ and then becomes narrow along with an increasing rotation angle, and eventually vanish for $30^\circ$. Furthermore, the first topological phase transition angle monotonically decreases with the relative lattice constant $a_0/R$, shown as the solid curve in panel (g). When ${a}_{0}/R$ reaches 2.6, the neighbouring atoms become tangent, which hinders designers to tune $a_0/R$ further. In short, the rule of thumb for setup parameters to achieve a topological phase indicates a range of ${a}_{0}/R=2.6$ to 3, when the rotation angle of the unit cell can be tuned to close and open the topological band gap resulting from band inversion. It is worth noting that for ${a}_{0}/R>3.0$ in the trivial phase $\mathrm{II}$, any angle of rotation will not cause phase switching though a topologically trivial band gap still occurs near $\Gamma$ point. Note that the rotation mechanism preserves the pseudo-TR of the system (cf. Sec.~$\mathrm{III}$ of Supplemental Material~\cite{SM}), and eases the method to generate the topological band gap without using magnetic experimental setups~\cite{2019Topological,Xu2010Quantum,2008Possible}.

The hallmark of topological bands is the scattering-free boundary states propagating on the interfaces between distinct topological matters~\cite{Xu2010Quantum,2019Topological,2008Possible}. To reveal the edge states explicitly in our QSHE system, we plot the projected band dispersion and electric field $\mathbf{E}_z$ in a ribbon-shaped supercell composed of two distinct PCs, respectively $\alpha=0^\circ$ and $\alpha=12^\circ$ in Fig.~3. As the dispersion diagram in Fig.~3(a) shows, a pair of edge modes (represented by red dot A and blue dot B) are observed within the bandgap. This pair of edge states is topologically protected by pseudo-TR symmetry, which can be identified as the spin-up and down modes. Actually, there is a tiny gap at the $\Gamma$ point in Fig. 3(a) [unnoticeable in the present scale] due to the symmetry of $C_6$ being damaged to a certain extent at the interface between the two crystals.   However, compared to the large size of the two crystals, $C_6$ symmetry could be taken as approximately kept and our topological properties maintained (cf. Sec.~$\mathrm{I V}$ of Supplemental Material~\cite{SM}). In left panel of (b) we plot the electric fields of the edge state for points A and B at $k_{x}$ is $\pm 0.02$ in units of $2\pi /a_{0}$ in (a). The field maps indicate the robust topologically protected edge states along the interface. In the two zoom-ins in the right panel of Fig.~3(b), the two circular arrows (red and blue for clockwise and counterclockwise rotation) indicate time-averaged power flow directions of $\mathbf{S}=\operatorname{Re}\left[\mathbf{E} \times \mathbf{H}^{*}\right] / 2$ for the edge states duo~\cite{Litvin2011Poynting}. It demonstrates the helical edge states under spin-momentum locking explicitly. 

\begin{figure}
\includegraphics[width=0.5\textwidth]{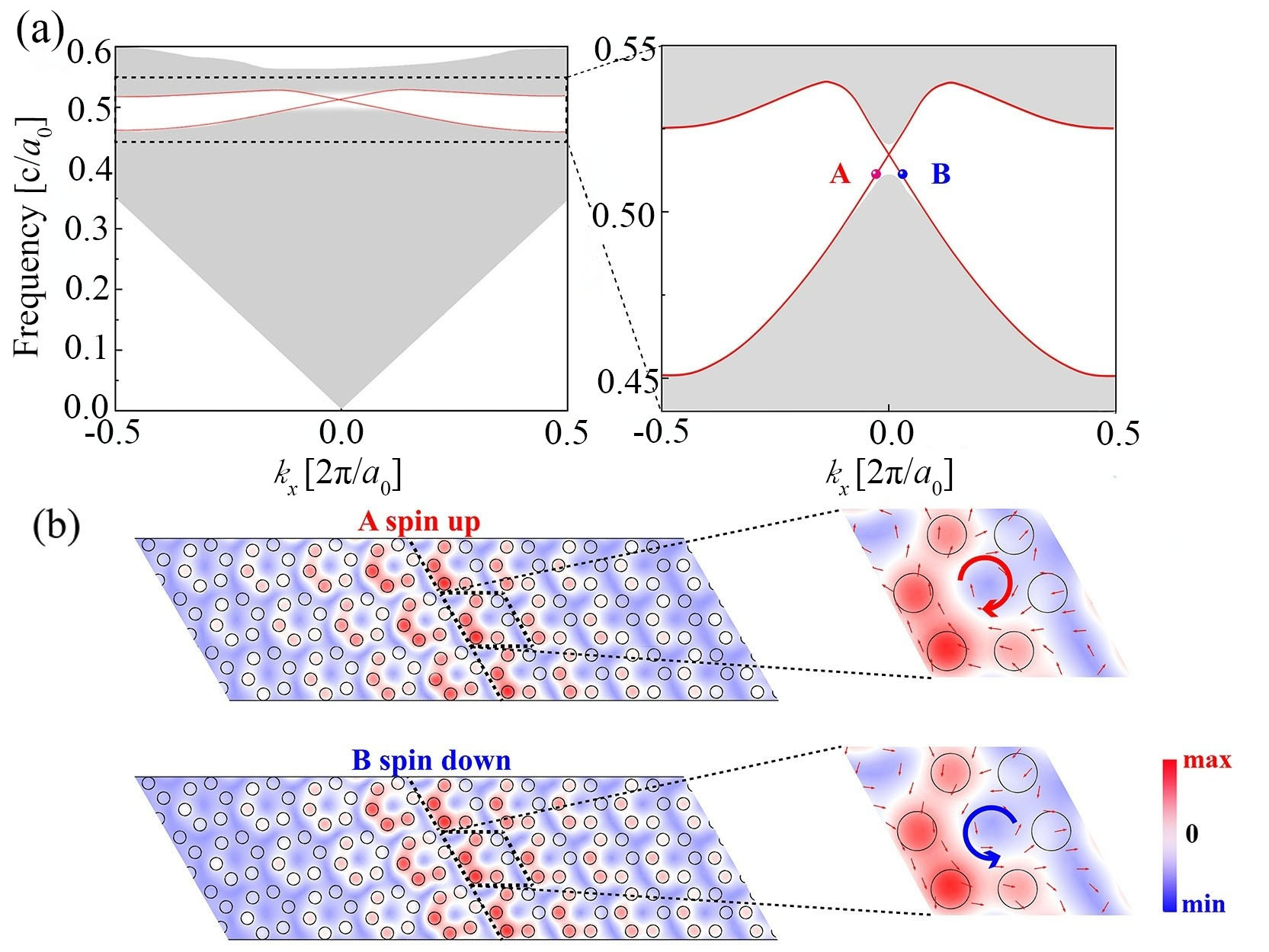} 

\caption{\label{fig:epsart} (a) Projected band diagram of a one-dimensional supercell, composed of both a topologically nontrivial phase $\mathrm{I}$ ($\alpha=0^\circ$) and topologically trivial phase region $\mathrm{II}$ ($\alpha=12^\circ$). A (red dot) and B (blue dot) correspond to the clockwise spiral and counterclockwise helical edge states,  respectively. (b) Left panel: Electric field distribution $\mathbf{E}_z(x, y)$ around the edge. Right panel: Zoom-in for electric field $\mathbf{E}_{z}$. The arrows represent the average Poynting vector directions and the magnitude corresponding to points A and B for $k_{x}$ is $\pm 0.02$, in units of $2\pi /a_{0}$ and circular thick arrows in red and blue are guides for the eyes.}
\end{figure}

To further validate the robust one-way propagation of topological helical edge states subject to PC defects~\cite{Wang2020Valley, Miyake2016Design}, we use the helicity feature, i.e. the direction of the pseudo-spins, of the edge states to control the propagation of electromagnetic waves. The pseudo-spin-up (pseudo-spin-down) mode is selectively excited by using a positive (negative) circular polarization excitation source $S_+$ ($S_-$)~\cite{2008Possible}. On the interface of two kinds of photonic crystals with distinct topological phases, we observe the EM propagation mostly along the edge. In the waveguide design, we make use of the wider common bandgap to excite the topological edge state selectively. 
For the topological nontrivial ($\rm{I}$) part the parameters are kept the same as Fig.~3, while for the trivial phase ($\rm{II}$) part, parameters $a_{0}$/R =3, $\alpha=12^\circ$ are chosen with other parameters unchanged. When the working frequency is chosen as 0.52 in units of $c/a_0$, the in-line waveguide excites EM waves by the pseudo-spin source $S_+$. Simulation shows that these edge states can transmit EM waves well in the selected rightward direction as shown in Fig.~4(a). To verify the defect immunity characteristics of the topological edge state, an in-line waveguide with defects is designed as shown in Fig.~4(b). Simulation also verifies that EM waves are still able to transmit around the disordered region to localize at the topological interface. Furthermore, to validate the backscattering immunity characteristics of the edge states, a Z-shaped waveguide with sharp bends is shown in Fig.~4(c). The EM wave excited from $S_+$ propagating along the Z-shaped interface between the topological distinct phases ($\rm{II}$/$\rm{I}$) can travel around the geometric bends without significant loss. In order to represent the energy flow in the waveguide excited by different chiral excitation sources, the Poynting vector distribution of the in-line waveguide and the defective waveguide are plotted in panels (d-e), which is distributed in a one-way transverse vortex during its propagation. By calculating the transmission efficiency of the three waveguides in panels (a-c), as shown in Fig.~4(f), we confirm that the energy transmission of the waveguides around the band gaps $\rm{I}$ and $\rm{II}$ has negligible loss. In order to further verify the topological feature of the edge-state propagation, two boundary types of waveguides are designed, in which the zigzag boundary is used in Fig.~3 and the armchair-type boundary state also demonstrates backscattering suppression and null-interference properties (cf. Sec.~$\mathrm{V}$ of Supplemental Material~\cite{SM}). Our results confirm that the boundary states induced by our rotation-induced phase transition are a true analogue of QSHE states, which are robust one-way reflection-less travelling eigenwaves against defects~\cite{Xu2010Quantum, 2008Possible}. 
\begin{figure}
\includegraphics[width=0.5\textwidth]{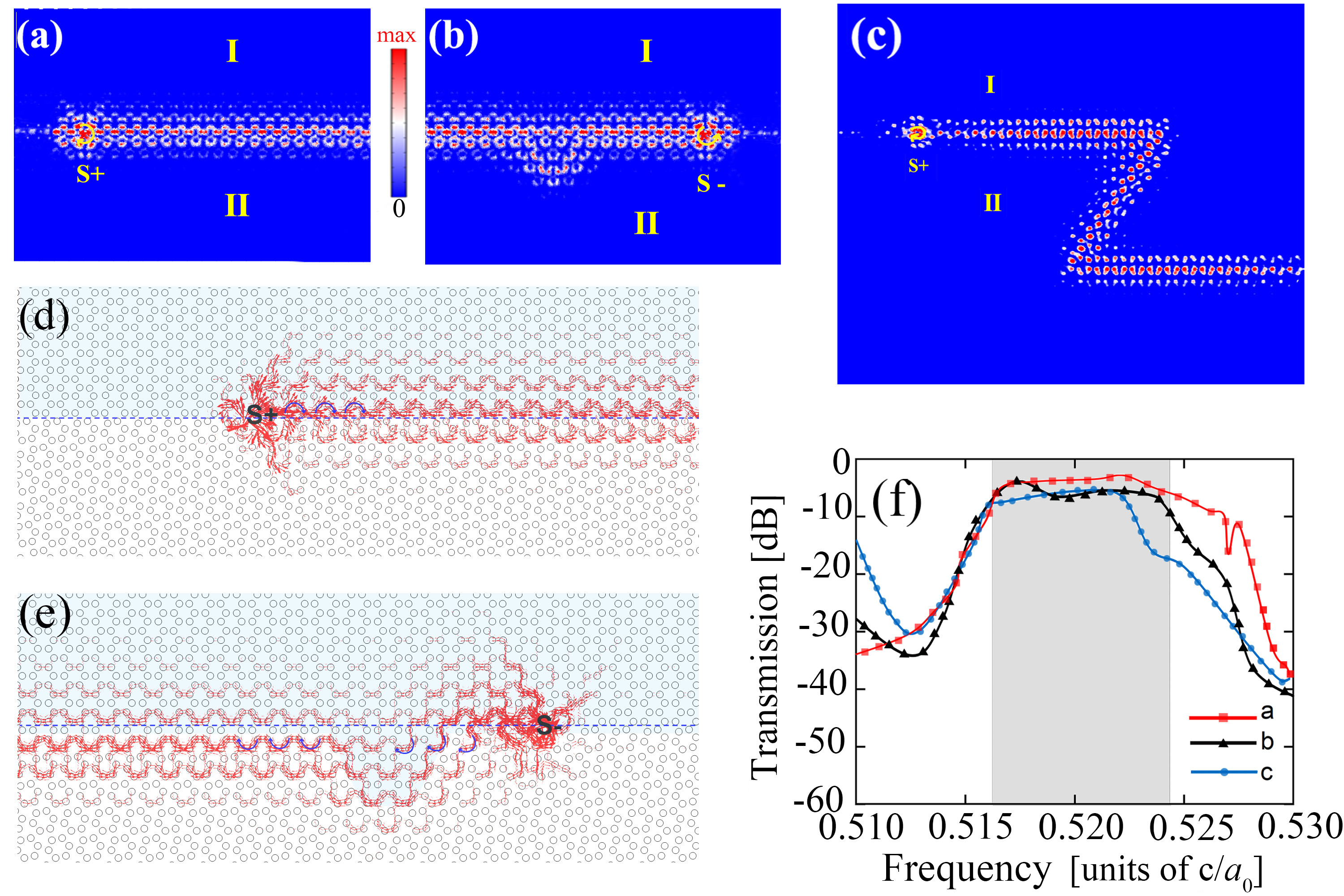}
\caption{\label{fig:epsart} A defect waveguide along the interface between two types of PCs ($\mathrm{I}:\alpha=0^\circ$ and $\mathrm{II}:\alpha=12^\circ, {a}_{0}/R =3$). The positive (negative) circular polarization excitation source $S_+ (S_-)$ represents the pseudo-spin-up (down) mode with frequency 0.52 in units of $c / a_0$. (a) Electric field of the edge state is excited by pseudo-spin-up mode $S_+$, showing that the EM wave propagates unidirectionally to the right. (b) In presence of defects, the crystal lattice appears disordered on the interface between $\mathrm{I}$ and $\mathrm{II}$ and the spin-down mode is still able to excite reflectionless transmission. (c) On an interface with a sharp angle, EM waves can also steer around geometric bends without backscattering. (d) Poynting vector distribution of an in-line waveguide excited by $S_+$ source. (e) Poynting vector distribution of  $S_-$  source excited in a bent waveguide. (f) Transmission spectra measured around band gaps in gray $\mathrm{I}$ and $\mathrm{II}$ [lines with red squares, black triangles and blue circles respectively represent the corresponding waveguides in (a-c)].}
\end{figure}

\section{\label{sec:level3}CONCLUSIONS}
In summary, based on the pure-dielectric Kekul\'e lattice, we use the synchronized rotation of the unit cells to induce a topological phase transition on a PC platform. The pseudo-TR symmetry is constructed based on the $C_6$ symmetry of the Kekul\'e lattice in design, and the topological phase transition in the rotation mechanism maintains the pseudo-TR inversion. Our model exploits the rotational freedom of the pillars to break the mirror symmetry of the crystal lattice, thereby reducing the manufacturing requirements to manipulate the Dirac points. The synchronized rotation of unit cells directly opens the bandgap, providing an additional degree of freedom for the generation of a topological gap without magnetic experimental setups. Electromagnetic wave simulation verifies that topological edge states emerge within a band gap due to the synchronized rotation of unit cells. Other than the already-known platforms~\cite{Chang2013Experimental,G2013Optical,Chen2014Experimental,Khanikaev2015Topologically,Fang2012Realizing}, we hope that this work provides a design possibility for scrutinizing optical QSHE systems. 


\begin{acknowledgments}
We thank Xu Donghui and Chen Menglin for helpful discussion to improve our understanding on topological photonic crystals. We are supported by the fundamental Research Funds for
the Central University of China [CCNU18JCXK02,
CCNU18GF006, CCNU16A02016, CCNU19TS073]; the
open fund of Guangxi Key Laboratory of Wireless Wideband communication and Signal Processing [GXKL06190202]; the open fund of China Ship Development
and Design Centre [XM0120190196]; National Natural Science Foundation of China [NSFC11804087]; and in part
by the Beijing Orient Institute of Measurement and Test
Electrostatic Research Foundation of Liu Shanghe Academicians
and Experts Workstation [BOIMTLSHJD20181002]. 
\end{acknowledgments}



\providecommand{\noopsort}[1]{}\providecommand{\singleletter}[1]{#1}%
%

%



\pagebreak

\onecolumngrid
\begin{center}
  \textbf{\large Supplemental material: Topological edge states of Kekul\'e-type photonic crystals induced by a synchronized rotation of unit cells}\\[.2cm]
 R. Zhou$^{1}$, H. Lin$^{1, *}$, Y. Liu$^{2, \dagger}$, X. Shi$^{1}$, R. Tang$^{3}$, Y. Wu$^{1}$ and Z. Yu$^{1}$
 \\[.1cm]
  {\itshape ${}^1$College of Physics Science and Technology, Central China Normal University, Wuhan 430079, Hubei Province\\
  ${}^2$School of Physics and Electronic Sciences, Hubei University, Wuhan 430062, Hubei Province\\
  ${}^3$Institute of Space Science and Technology, Nanchang University, Nanchang 330021, Jiangxi Province}\\
  ${}^*$Electronic address: linhai@mail.ccnu.edu.cn; \\
  ${}^\dagger$Electronic address: yangjie@hubu.edu.cn\\

(Dated: \today)\\[1cm]
\end{center}

\setcounter{equation}{0}
\setcounter{figure}{0}
\setcounter{table}{0}
\setcounter{page}{1}
\renewcommand{\theequation}{S\arabic{equation}}
\renewcommand{\thefigure}{S\arabic{figure}}
\renewcommand{\bibnumfmt}[1]{[S#1]}
\renewcommand{\citenumfont}[1]{S#1}

\section{\label{sec:level1}
Photonic pseudo-spin related to $C_6$ symmetry}

The entire crystal structure maintains $C_6$ symmetry. According to the group theory, any eigenstate of the crystal at $\Gamma$ point corresponds to an irreducible representation in the $C_6$ point group. There are two irreducible representations called $\mathbf{E}_{1}$ and $\mathbf{E}_{2}$  irreducible representations~\cite{Xu2010Quantum} in two-dimension. The eigenstate of $\mathbf{E}_{1}$  has the opposite parity for the mirror operation of ${x}$-axis and ${y}$-axis, and the eigenstate corresponding to $\mathbf{E}_{2}$ has the same parity for the mirroring operation of ${x}$-axis and ${y}$-axis. In other words, $\mathbf{E}_{1}$  and $\mathbf{E}_{2}$ eigenstates respectively carry odd parity and even parity under the space inversion operation. In addition,  $\mathbf{E}_{1}$  and $\mathbf{E}_{2}$  eigenstates are both double degenerate, whose basis functions are ${(x, y)}$ and $({x^{2}-y^{2}},{2xy})$. They have the same symmetry with ($p_{x},p_{y})$,($d_{x y},d_{x^{2}-y^{2}})$ states~\cite{ Xu2010Quantum,Wu2015Scheme}.

With ${(x, y)}$ as the basis function, the matrix corresponding to $\mathbf{E}_{1}$ rotation of $\pi / 3$ is: 
\begin{equation}
D_{E_{1}}\left(\boldsymbol{C}_{6}\right)\left(\begin{array}{l}
\mathrm{p}_{x} \\
\mathrm{p}_{y}
\end{array}\right)=\left(\begin{array}{cc}
1 / 2 & -\sqrt{3} / 2 \\
\sqrt{3} / 2 & 1 / 2
\end{array}\right)\left(\begin{array}{l}
\mathrm{p}_{x} \\
\mathrm{p}_{y}
\end{array}\right). 
\end{equation}
Define a unitary operator:
\begin{equation}
\boldsymbol{U}=\frac{1}{\sqrt{3}}\left[\boldsymbol{D}_{E_{1}}\left(\boldsymbol{C}_{6}\right)+\boldsymbol{D}_{E_{1}}\left(\boldsymbol{C}_{6}^{2}\right)\right]=-\mathrm{i} \sigma_{y}, 
\end{equation}
where $\sigma_{y}$ is the second Pauli operator.
With ${x, y} ({x^{2}-y^{2}})$ as the basis functions, the matrix corresponding to $\mathbf{E}_{2}$ is:
\begin{equation}
D_{E_{2}}\left(\boldsymbol{C}_{6}\right)\left(\begin{array}{c}
\mathrm{d}_{x^{2}-y^{2}} \\
\mathrm{~d}_{x y}
\end{array}\right)=\left(\begin{array}{cc}
-1 / 2 & -\sqrt{3} / 2 \\
\sqrt{3} / 2 & -1 / 2
\end{array}\right)\left(\begin{array}{c}
\mathrm{d}_{x^{2}-y^{2}} \\
\mathrm{~d}_{x y}
\end{array}\right).
\end{equation}
Correspondence operator:
\begin{equation}
\boldsymbol{U}=\frac{1}{\sqrt{3}}\left[\boldsymbol{D}_{E_{1}}\left(\boldsymbol{C}_{6}\right)-\boldsymbol{D}_{E_{1}}\left(\boldsymbol{C}_{6}^{2}\right)\right]=-\mathrm{i} \sigma_{y}
\end{equation}
$\boldsymbol{U}$ and the complex conjugate operator $\boldsymbol{K} $ by combining form an anti-unitary operator $\mathrm{T}_{\mathrm{S}}:\mathrm{T}_{\mathrm{S}}=\boldsymbol{U K}$.
as can be verified:
\begin{equation}
\begin{gathered}
T_{\mathrm{s}}^{2}\left(\begin{array}{l}
\mathrm{p}_{x} \\
\mathrm{p}_{y}
\end{array}\right)=T_{\mathrm{s}}\left(\begin{array}{c}
-\mathrm{p}_{y} \\
\mathrm{p}_{x}
\end{array}\right)=-\left(\begin{array}{c}
\mathrm{p}_{x} \\
\mathrm{p}_{y}
\end{array}\right), \\
T_{\mathrm{s}}^{2}\left(\begin{array}{c}
\mathrm{d}_{x^{2}-y^{2}} \\
\mathrm{~d}_{x y}
\end{array}\right)=T_{\mathrm{s}}\left(\begin{array}{c}
-\mathrm{d}_{x y} \\
\mathrm{~d}_{x^{2}-y^{2}}
\end{array}\right)=-\left(\begin{array}{c}
\mathrm{d}_{x^{2}-y^{2}} \\
\mathrm{~d}_{x y}
\end{array}\right) .
\end{gathered}
\end{equation}
It can be seen from the above analysis that in the designed photonic crystal system, $T_{s}^{2}=-1$ for both $\mathbf{E}_{1}$ and $\mathbf{E}_{2}$ modes. This is analogous to the time-reversal symmetry (TRS) in the electronic system, which ensures that the photonic crystal has Kramers degeneracy at $\Gamma$ point. Therefore, we exploit $T_{s}$ the pseudo-TR operator of photonic crystals in our manuscript. 

In a photonic crystal with $C_6$ symmetry, the pseudo-spin state is defined as:
 \begin{equation}
\begin{aligned}
&\mathrm{p}_{\pm}=\left(\mathrm{p}_{x} \pm \mathrm{i} \mathrm{p}_{y}\right) / \sqrt{2} \\
&\mathrm{~d}_{\pm}=\left(\mathrm{d}_{x^{2}-y^{2}} \pm \mathrm{id}_{x y}\right) / \sqrt{2} .
\end{aligned}
\end{equation}
The unitary transformation matrix between the representation for $[p_{+}, p_{-}]$ as the basis functions and that for $p_{x} (p_{y})$ is:
\begin{equation}
S=\left(\begin{array}{cc}
1 / \sqrt{2} & 1 / \sqrt{2} \\
i / \sqrt{2} & -i / \sqrt{2}
\end{array}\right)
\end{equation}
Therefore, when $p_{x} (p_{y})$ is used as the basis functions, the operator defined by equation (4) becomes:
\begin{equation}
\boldsymbol{U}^{\prime}=\boldsymbol{S}^{+} \boldsymbol{U} \boldsymbol{S}=\left(\begin{array}{cc}
-\mathrm{i} & 0 \\
0 & \mathrm{i}
\end{array}\right)
\end{equation}
The corresponding pseudo-TR operator is $T_{\mathrm{s}}^{\prime}=U^{\prime} K$:
\begin{equation}
T_{\mathrm{s}}^{\prime} \mathrm{p}_{\pm}=\mp \mathrm{ip}_{\pm}, T_{\mathrm{s}}^{\prime 2} \mathrm{p}_{\pm}=-\mathrm{p}_{\pm}
\end{equation}
It can be seen from Eq. (S9) that the wave function $[p_{+}, p_{-}]$ corresponds to the pseudo-spin irreducible representing $\mathbf{E}_{1}$. Under the pseudo-TR operation, $p_{+}$ represents the pseudo-spin-up state, and $p_{-}$ represents the pseudo-spin-down state. The same conclusion is drawn for $[d_{+}, d_{-}]$, which means the pseudo-spin-up and pseudo-spin-down states corresponding to $\mathbf{E}_{2}$. We note that our pseduo-spins could also be defined differently from modal hybridization, e.g. from circular polarization instead~\cite{2018Elastic}.

\section{\label{sec:level2} The $\mathbf{k}\cdot\mathbf{p}$ theory}
In this section, we discuss the interaction between the energy bands (${p}$ state and ${d}$ state) of photons based on the $\mathbf{k}\cdot \mathbf{p}$ perturbation theory~\cite{2016Accidental}. First Maxwell equation in TM mode can be written as:
\begin{equation}
\left[\frac{1}{\varepsilon(\mathbf{r})} \nabla \times \nabla \times\right] E_{z}(\mathbf{r}) \hat{z}=\frac{\omega^{2}}{c^{2}} E_{z}(\mathbf{r}) \hat{z},
\end{equation}
where $\varepsilon({r})$ is the dielectric parameter and $c$ the speed of light. When $\varepsilon({r})$ is periodic, Bloch theorem applies in form of EM waves.

In order to understand the topological properties of the two crystal band gaps, we obtain the effective Hamiltonian near $\Gamma$ point according to the $\mathbf{k}\cdot \mathbf{p}$ perturbation theory, and by comparing the Bernevig-Hughes-Zhang (BHZ) model~\cite{Bernevig2006} in electronic system to understand the corresponding topological features of the system. 

The four eigenstates near $\Gamma$ point are:
\begin{equation}
\begin{gathered}
\Gamma_{\alpha}(\alpha=1,2,3,4): \Gamma_{1}=\mathrm{p}_{x}=|x\rangle, \Gamma_{2}=\mathrm{p}_{y}=|y\rangle \\
\Gamma_{3}=\mathrm{d}_{x^{2}-y^{2}}=\left|x^{2}-y^{2}\right\rangle, \Gamma_{4}=\mathrm{d}_{x y}=|x y\rangle
\end{gathered}
\end{equation}
According to the $\mathbf{k}\cdot \mathbf{p}$ perturbation theory, 
\begin{equation}
\mathcal{H}(\mathrm{k})=\mathcal{H}_{0}+\mathcal{H}^{\prime}, 
\end{equation}
where:
\begin{equation}
\mathcal{H}_{0}=\left(\begin{array}{cccc}
\epsilon_{p}^{0} & & & \\
& \epsilon_{p}^{0} & & \\
& & \epsilon_{d}^{0} & \\
& & & \epsilon_{d}^{0}
\end{array}\right). 
\end{equation}
 
\begin{equation}
\mathcal{H}^{\prime}=\left(\begin{array}{cccc}
G k_{x}^{2}+F k_{y}^{2} & N k_{x} k_{y} & A k_{x} & A k_{y} \\
N k_{x} k_{y} & F k_{x}^{2}+G k_{y}^{2} & -A k_{y} & A k_{x} \\
A^{*} k_{x} & -A^{*} k_{y} & -G k_{x}^{2}-F k_{y}^{2} & -N k_{x} k_{y} \\
A^{*} k_{y} & A^{*} k_{x} & -N k_{x} k_{y} & -F k_{x}^{2}-G k_{y}^{2}
\end{array}\right). 
\end{equation}

There $P_{x}=\partial / \partial x$, $P_{y}=\partial / \partial y$, $F=\left|\left\langle x\left|P_{y}\right| 2 x y\right\rangle\right|^{2} /\left(E_{1}-E_{4}\right)$, $G=\left|\left\langle x\left|P_{x}\right| x^{2}-y^{2}\right\rangle\right|^{2} /\left(E_{1}-E_{3}\right)$.

According to the $\mathbf{k}\cdot \mathbf{p}$ theory, effective photonic Hamiltonian under the representation of $\left[\begin{array}{llll} p_{+}, & d_{+}, & p_{-}, & d_{-}\end{array}\right]^{T}$ is: 
\begin{equation}
\boldsymbol{H}(k)=\left(\begin{array}{cccc}
M-B k^{2} & A k_{+} & 0 & 0 \\
A^{*} k_{-} & -M+B k^{2} & 0 & 0 \\
0 & 0 & M-B k^{2} & A k_{-} \\
0 & 0 & A^{*} k_{+} & -M+B k^{2}
\end{array}\right)
\end{equation}
In Eq.~(S15), $k_{\pm}=k_{x} \pm \mathrm{i} k_{y}$, $A$ is the coupling coefficient between p-and d-states, $M=\left(\epsilon_{d}^{0}-\epsilon_{p}^{0}\right) / 2$, $B=-(F+G)/2-D$, $D=\left(\epsilon_{d}^{0}+\epsilon_{p}^{0}\right) / 2$. For this proposal, the pseudo-spins are emulated by the orbital angular momenta, which serve as the photonic analog of the electronic Kramers pairs, and the effective Hamiltonian for photons can be directly mapped to BHZ model for the quantum spin Hall insulators of electrons. 
Based on Hamiltonian in Eq.~(S15), Chern numbers for two pseudo-spins can be written as \cite{Wu2015Scheme}:
\begin{equation}
\mathcal{C}_{\pm}=\frac{1}{2 \pi i} \int d^{2} \mathrm{k}\left(\frac{\partial \mathcal{A}_{y}^{\pm}}{\partial k_{x}}-\frac{\partial \mathcal{A}_{x}^{\pm}}{\partial k_{y}}\right)
\end{equation}
Where $\overrightarrow{\mathcal{A}^{\pm}}=\left(\mathcal{A}_{x}^{\pm}, \mathcal{A}_{y}^{\pm}\right)=\left\langle\psi_{\mathbf{k}}^{\pm}\left|\nabla_{\mathbf{k}}\right| \psi_{\mathbf{k}}^{\pm}\right\rangle$ with $\psi_{\mathbf{k}}^{\pm}$ the two-component eigenstates of $\mathcal{H}^{\pm}$. For a synchronized rotation of unit cells, the p bands are lying well up the d
bands (see Fig. 2(a) in the main text),  by integrating Berry curvatures over the
first Brillouzin zone respectively, that the Chern numbers
for pseudo spin-up and -down channels are $\mathcal{C}_{\pm}=\pm 1$. In the case that a band inversion of d and p orbitals happens at $\Gamma$ point (see Fig. 2(c) in the main text), for which the
Chern numbers are zero indicating that the system takes
a trivial state.

\section{Unit cells rotation with pseudo-TR symmetry}
\begin{figure}
\includegraphics[width=0.5\textwidth]{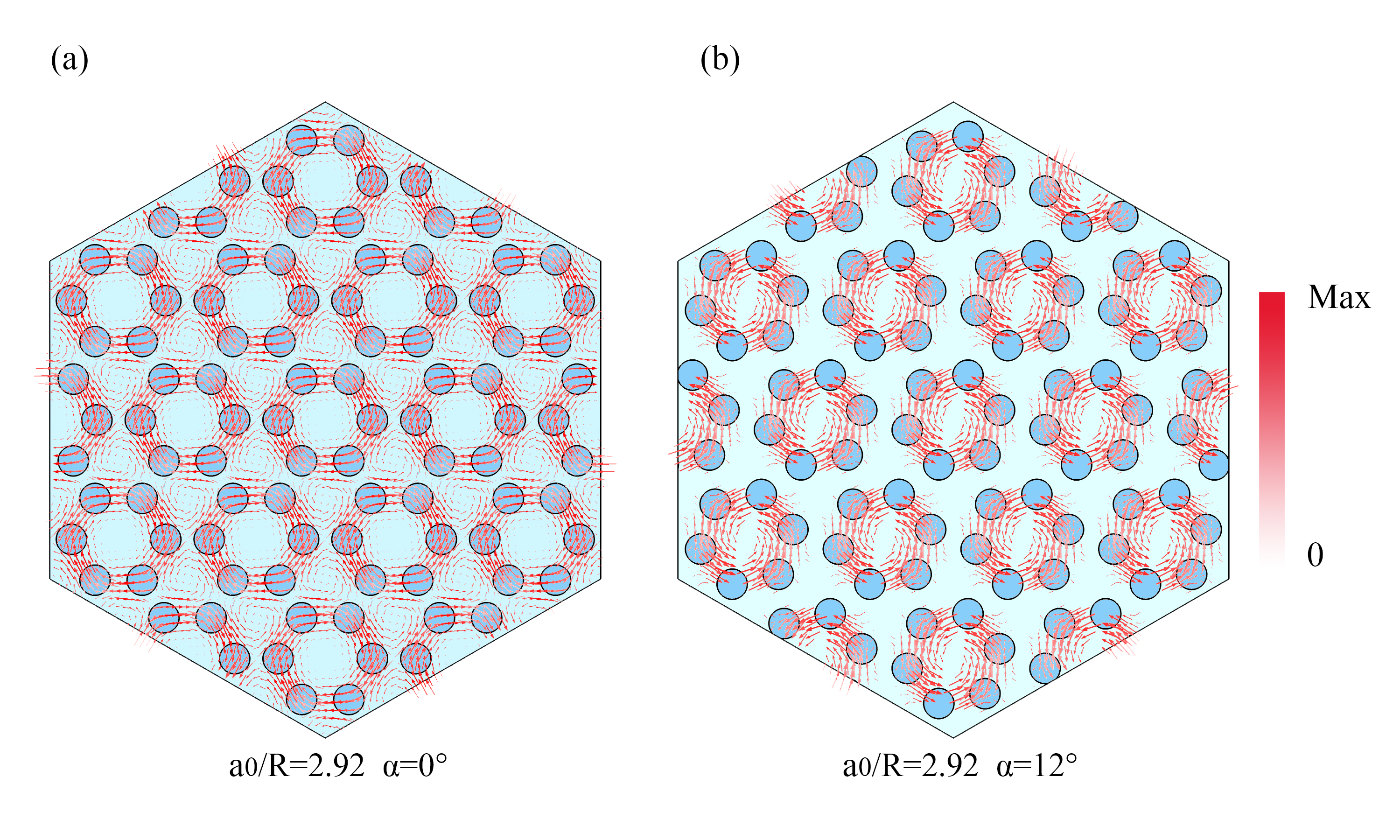} 
\caption{\label{fig:epsart}Spatial distribution of Poynting vector under the photonic band gap of $\Gamma$ point. (a) ${a}_{0}/R=2.92$,$\alpha=0^\circ$, Poynting vector rotates between the unit cells. (b) ${a}_{0}/R=2.92$,$\alpha=12^\circ$, Poynting vector rotates around inside the unit cells.}
\end{figure}
 
In order to observe the topological phases before and after the internal rotation unit cells, we check the spatial distribution of the pseudo-spin specific Poynting vector when the rotation degree are $\alpha=0^\circ$ and $\alpha=12^\circ$. As shown in Fig.~S1(a), before rotation $\alpha=0^\circ$ the Poynting vector rotates between unit cells due to the topological band gap~\cite{Wu2015Scheme}. When the unit cells rotates $\alpha=12^\circ$ as shown in Fig.~S1(b), Poynting vector only rotates around inside the unit cells, in a topologically-trivial state. The internal rotation of the unit cells only breaks the mirror symmetry of the spatial point group, but does not change the $C_6$ symmetry of the point group. By observing the real space distribution of the Poynting vector down the photonic band gap near $\Gamma$ point, it is found that the chirality of the Poynting vector corresponds to the pseudo-spin, resulting from the pseudo-TR symmetry of the spatial point group.

The internal rotation of the unit cells breaks the mirror symmetry of the system, so that the electromagnetic energy distribution of the system gradually changes during synchronized rotation from a topologically non-trivial state to a topologically trivial state. This transition is accompanied by the Kramer's degenerence at near $\Gamma$ point. When the photonic system maintains the symmetry of pseudo-time reversal, an accidental node (quadruple degeneracy point) appears at $\Gamma$ when rotated by a certain angle. The accidental node carries a different dispersion and topological properties (Berry phase), which is the critical state between phases~\cite{2011Dirac}.

\section{Spatially-deformed topological interface with moderate breaking of pseudo-time reversal symmetry}
By cutting the honeycomb lattice structure in different directions, two typical boundaries can be obtained: zigzag and armchair type. Therefore, putting together photonic crystals with different topological states will form two typical interfaces. In order to confirm the existence of topological boundary states, we calculate the projection bands of the supercells of two boundary states composed of topologically trivial (${a}_{0}/R=2.92$, $\alpha=12^\circ$) and topologically non-trivial (${a}_{0}/R=2.92$,$\alpha=0^\circ$) as shown in Fig.~S2(a-b). We clearly observe the bulk-edge correspondence by increasing the period length of the supercell in x direction as shown in Fig.~S2(c-d).
 
\begin{figure}
\includegraphics[width=0.5\textwidth]{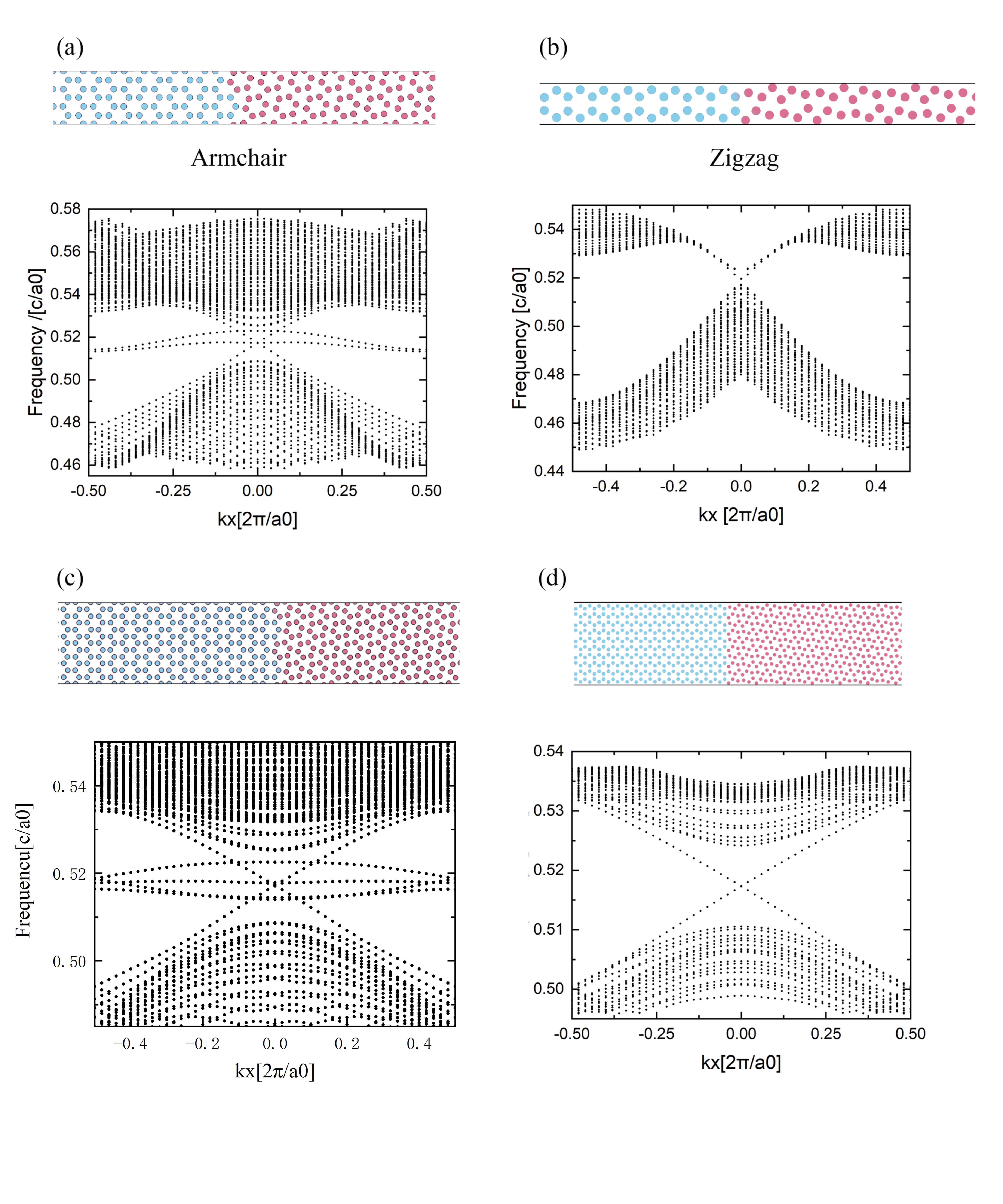} 
\caption{\label{figS2} Projection bands of two typical interfaces composed of topological  non-trivial phases  $\mathrm{I}$ (${a}_{0}/R=2.92$,$\alpha=0^\circ$) and topologically trivial phases  $\mathrm{ II }$ (${a}_{0}/R=2.92$,$\alpha=12^\circ$). (a) One-unit thick armchair interface, the two bands in the middle of the band gap represent the boundary states of the upper and lower boundaries. (b) One-unit thick zigzag interface. (c) Two-unit thick armchair interface. (d) Five-unit thick zigzag interface.}
\end{figure}

It can be seen from Fig.~S2(a)-(b) that the two boundary states retain the symmetry of pseudo-time inversion, and the linearly-crossed edge states with opposite directions represent the pseudo-spin up and down state. By comparing Fig.~S2(a) and (c), we find that the dispersions near $0.52c/a_0$ are actually doubly degenerate, which can be identified by increasing the period length along x direction.

\section{Unidirectional transmission of topological edge states}
In order to single out a single pseudo-spin for the topological edge states on crystal interfaces, a chiral point source composed of positive and negative circularly polarized excitation sources $S_+ (S_-)$ are used. The chiral point source can excite the pseudo-spin up/down state. A zigzag-type boundary is already used to form a waveguide Verified in maintext Fig.4. Here we use an $S_+ (S_-)$ point source to excite an armchair-type boundary state in a waveguide.
 \begin{figure}
\includegraphics[width=0.5\textwidth]{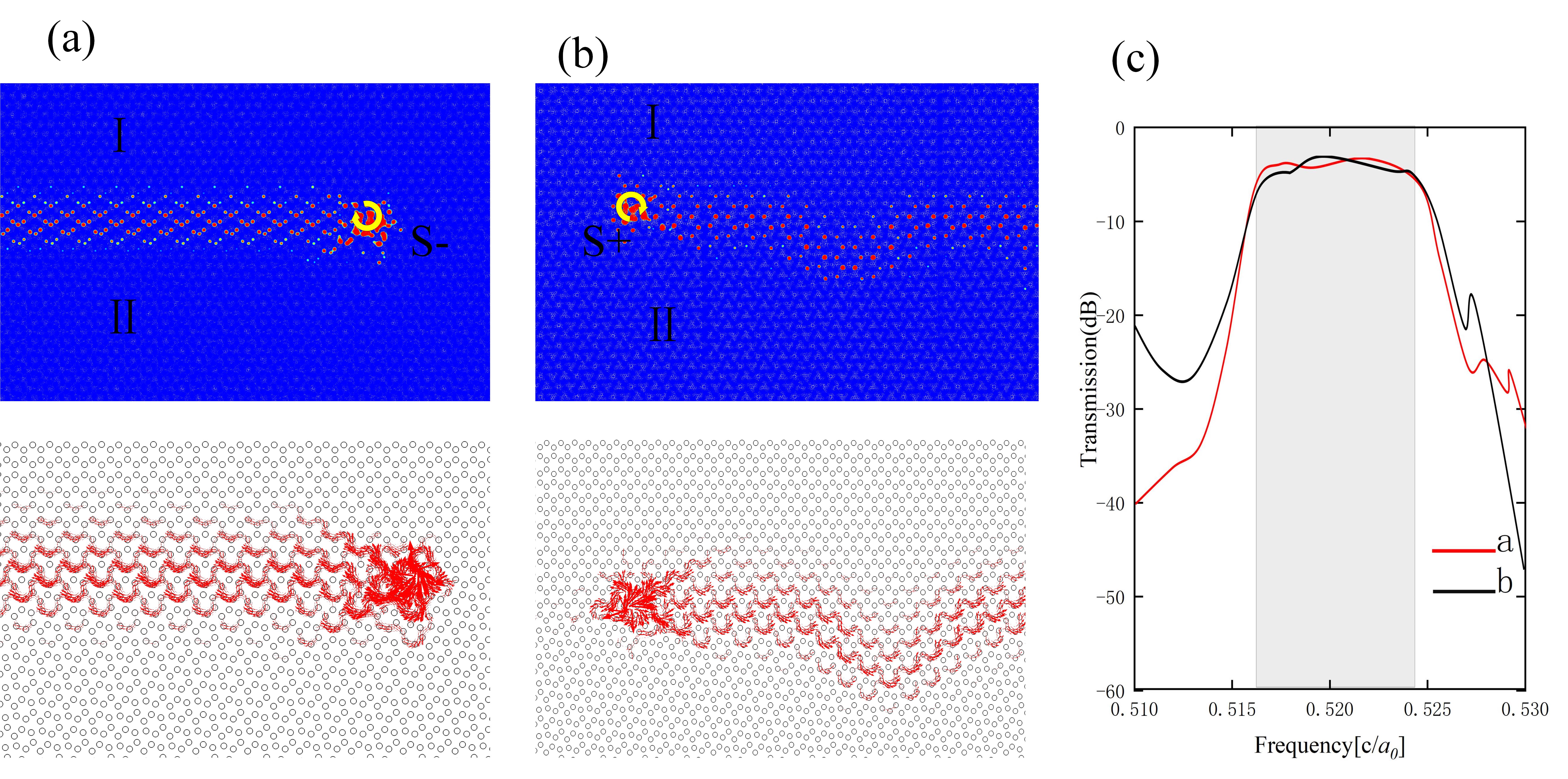} 
\caption{\label{fig:sepsart}Unidirectional transmission of the pseudos-spin along the armchair-typeinterface between phase $\mathrm{I}$ (${a}_{0}/R=2.92$,$\alpha=0^\circ$) and $\mathrm{II}$ (${a}_{0}/R=3$,$\alpha=12^\circ$ ). (a) The $S_-$ point source is used to excite the in-line waveguide, where the energy flow vector of the lower panel is partially enlarged. (b) The $S_+$ point source is used to excite the disordered defect waveguide. (c) Measured transmission spectra around the band gap (gray shaded area). The red and black lines represent transmittance of the in-line waveguide in Fig.S3(a) and (b) respectively.}
\end{figure}
In order to further verify the topological characteristics of our design, we designed two boundary types of waveguides. Fig. S3(a)-(b) show that the armchair-type boundary state is able to propagate without backscattering. By measuring the transmission spectra around the band gap of the two waveguides as shown in Fig. S3(c), we find that the topologically protected edge state is also highly transmissive.



\providecommand{\noopsort}[1]{}\providecommand{\singleletter}[1]{#1}%

\end{document}